\documentclass[twoside,11pt]{article}

\usepackage{jmlr2e}

\firstpageno{1}

\usepackage{multirow}

\usepackage{caption}
\usepackage{subcaption}

\usepackage{datatool}
\newcommand{\sortitem}[1]{%
  \DTLnewrow{list}
  \DTLnewdbentry{list}{description}{#1}
}
\newenvironment{sortedlist}{%
  \DTLifdbexists{list}{\DTLcleardb{list}}{\DTLnewdb{list}}
}{%
  \DTLsort{description}{list}
  \begin{itemize}%
    \DTLforeach*{list}{\theDesc=description}{%
      \item \theDesc}
  \end{itemize}%
}

\begin{document}
\includegraphics[width = .37\linewidth]{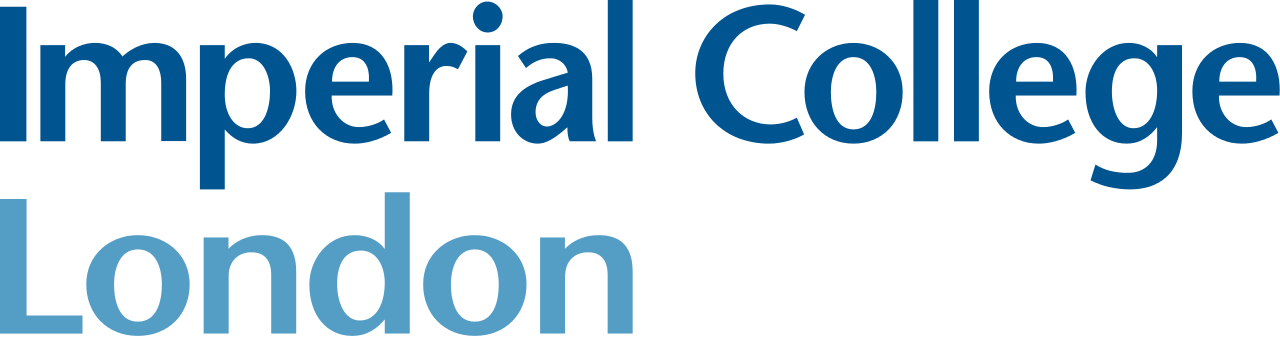}
\vspace*{.2in}

\title{Deep Convolutional Neural Networks to Diagnose COVID-19 and other Pneumonia Diseases from Posteroanterior Chest X-Rays}

\author{\name Pierre G. B. Moutounet-Cartan \email pierre.moutounet-cartan17@imperial.ac.uk \\
       \addr Department of Mathematics\\
       Imperial College London\\
       London, SW7 2AZ, United Kingdom}

\maketitle

\begin{abstract}
The article explores different deep convolutional neural network architectures trained and tested on posteroanterior chest X-rays of 327 patients who are healthy (152 patients), diagnosed with COVID-19\footnote{Referring to the coronavirus originating from Wuhan, mainland China, and discovered in 2019.} (125), and other types of pneumonia (48), of which Severe Acute Respiratory Syndrome (16), Streptococcus (13), Klebsiella (1), Legionellosis (2), Pneumocystis Jiroveci Pneumonia (13), Acute Respiratory Distress Syndrome not caused by COVID-19 (4), and Chlamydia Pneumoniae (1). In particular, this paper looks at the deep convolutional neural networks VGG16 and VGG19 \citep{simonyan:2015}, InceptionResNetV2 and InceptionV3 \citep{szegedy:2016}, as well as Xception \citep{chollet:2017}, all followed by a flat multi-layer perceptron and a final 30\% drop-out.\\

The paper has found that the best performing network is VGG16 with a final $30\%$ drop-out trained over 3 classes (COVID-19, No Finding, Other Pneumonia). It has a cross-validated training accuracy of $93.9(\pm3.4)$\%, a COVID-19 sensitivity of $87.7(-1.9,+2)$\%, and a No Finding sensitivity of $96.8(\pm0.8)$\%. The respective cross-validated values on the testing set are $84.1(\pm13.5)$\%, $87.7(-1.9,2)$\%, and $96.8(\pm0.8)$\%.\footnote{The values in brackets are the differences between the estimated values for the measures listed and the left or right boundary of the 95\% confidence interval after Stratified 5-Fold cross-validation. The inverse cumulative distribution function value is taken from a $t$-distribution with 4 degrees of freedom.} The model optimizer was Adam \citep{kingma:2015} with a 0.0001 learning rate, and categorical cross-entropy loss. \\

It is hoped that, once this research will be put to practice in hospitals, healthcare professionals will be able in the medium to long-term to diagnosing through machine learning tools possible pneumonia, and if detected, whether it is linked to a COVID-19 infection, allowing the detection of new possible COVID-19 foyers after the end of possible "stop-and-go" lockdowns as expected by \citet{ferguson:2020} until a vaccine is found and widespread. Furthermore, in the short-term, it is hoped practitioners can compare the diagnosis from the deep convolutional neural networks with possible RT-PCR testing results, and if clashing, a Computed Tomography could be performed as they are more accurate in showing COVID-19 pneumonia \citep{ai:2020, wang:2020, xu:2020}.

\end{abstract}
\vspace*{.1in}


\tableofcontents

\section*{Abbreviations}

\paragraph{Listed by alphabetical order}
\small
\begin{sortedlist}
  \sortitem{COVID-19: Novel Coronavirus 2019 SARS-n-CoV-2}
  \sortitem{PA (X-ray): Posteroanterior X-ray}
  \sortitem{CT: Computed Tomography}
  \sortitem{RT-PCR (test): Reverse Transcription Polymerase Chain Reaction test}
  \sortitem{SARS: Severe Acute Respiratory Syndrome}
  \sortitem{MERS: Middle East Respiratory Syndrome}
  \sortitem{ARDS: Acute Respiratory Distress Syndrome}
  \sortitem{2D-Conv($m, M$): 2-dimensional convolution layer of $M$ channels of $m\times m$ kernel}
  \sortitem{MaxPool($d, s$): MaxPool layer of $d\times d$ pool and $s\times s$ stride}
  \sortitem{Layer($n$): flat perceptron layer made of $n$ neurons}
  \sortitem{DropOut($p$): drop-out between two perceptron layers with probability $p$}
  \sortitem{Output($f$): output from the neural network through some activation function $f$}
  \sortitem{AI: Artificial Intelligence}
\end{sortedlist}
\normalsize

\newpage
\section{Introduction, Aims, \& State-of-the-Art}

As the writing of this article, there is a significant outbreak of COVID-19, which can lead to pneumonia, visible to medical imaging methods such as X-rays or CT scans. This pneumonia has also been detected in RT-PCR COVID-19 positive patients who did not have any known underlying health conditions \citep{cheng:2020}. Epidemiology research has tried to find the effective reproduction number before and after lockdowns and school closures, as well as estimating the real number of people infected by COVID-19 through mathematical models \citep{ferguson:2020, flaxman:2020} or seroprevalence studies \citep{bendavid:2020}. In particular, \citet{flaxman:2020} estimated that up to March 28, 2020, between 1.2\% and 5.4\% of the population had been COVID-19 infected in the United Kingdom (i.e., between roughly 800,000 and 3,600,000 residents of the United Kingdom). Antibody seroprevalence studies showed that between 2.24\% and 3.37\% of the Santa Clara, CA population contracted COVID-19 up to April 4, 2020 \citep{bendavid:2020}, and around 21\% of the New York City, NY population contracted the disease according to the Governor of New York. So far, the most advanced clinical trials for a potential vaccine are located in the United States, the United Kingdom, and Germany, and such a vaccine could only be available in fall 2020 at the earliest for healthcare professionals. \\ 

\citet{wang:2020} have applied deep convolutional neural networks on 1,065 CT images from the lungs of pathogen-confirmed COVID-19 cases (325 patients) along with those previously diagnosed with typical viral pneumonia (740 patients) not provoked by coronavirus. The internal validation in that research achieved a total accuracy of 89.5\% with specificity of 88\% and sensitivity of 87\%, while the external testing data-set showed a total accuracy of 79.3\% with specificity of 83\% and sensitivity of 67\%. 54 patients part of the study had the first two nucleic acid test results for COVID-19 were negative, of which 46 were predicted as COVID-19 positive by the algorithm with the probability of 85.2\%, which could further prove the very low accuracy of such tests. Indeed, \citet{ai:2020} discovered that of the patients with negative RT-PCR results, 75\% had positive chest CT findings of which 48\% were considered as highly likely cases of COVID-19.\\

Similarly, \citet{xu:2020} found that the real time reverse RT-PCR detection of viral RNA from sputum or nasopharyngeal swab has a relatively low positive rate to determine COVID-19 positiveness. After training ResNet-like neural networks on CT images of patients showing COVID-19 (357 images), Other Viral Pneumonia (390), and No Finding (963), that research found respective sensitivities of 81.5\%, 75.4\%, and 97.8\%, with an overall accuracy on the benchmark data-set of 86.7\%. These are seemingly similar results to \citet{wang:2020}.\\

An attempt of training deep neural networks on PA chest X-rays, where some of which showed a COVID-19 pathology, was made by \citet{narin:2020} with other parts of the same data-set used in this research, although they only used the classes COVID-19 (50 images) and No Finding (50), meaning their neural networks could, in principle, categorize non-COVID-19 pneumoniae as such if presented by such X-ray. This explains the high accuracies and other measures that they have found for the InceptionResNetV2, InceptionV3, and ResNet50 architectures. Indeed, they reached an accuracy of 98\%, a COVID-19 sensitivity of 96\%, and a specificity value of 100\% for ResNet50 on the external testing data. These are the cross-validated values through standard 5-Fold. Furthermore, the data-set was relatively small.\\

There are two main goals to this research: (1) in the medium to long-term, being able to diagnose through machine learning tools possible pneumonia, and if detected, whether it is linked to a coronavirus infection or other, allowing the detection of new possible coronavirus foyers after the end of possible "stop-and-go" lockdowns as expected by \citet{ferguson:2020} until a vaccine is found and widespread. (2) Furthermore, in the short-term, it can allow practitioners to compare the diagnosis from the deep convolutional neural networks in this paper with possible RT-PCR testing results -- if clashing, a chest CT could be done as they are more accurate in showing COVID-19 pneumonia as shown by \citet{ai:2020, wang:2020, xu:2020}.\\

This research focuses on PA chest X-rays because, although they can be less accurate to diagnose COVID-19 than CT imaging, they are more common in hospitals and relatively cheaper to perform. Hence, training neural networks on PA chest X-rays could lead to greater generalization and application in hospitals around the world, and CT imaging would only be used for cases where patients are pathogen-confirmed COVID-19 through RT-PCR testing but the AI diagnoses No Finding, or when the AI diagnoses the patients as highly-likely COVID-19 but the RT-PCR tests are negative.

\newpage
\section{The Data}

The data has been collected from \citet{cohen:2020} and \citet{kermany:2020}. The first source \citep{cohen:2020} had predominantly data from ill patients, and therefore the paper partially uses the second data-set \citep{kermany:2020} which is made of healthy patients only (only partially otherwise there would be an over-representation). \\

All training X-rays have been resized to a shape 182$\times$182. After train-test split, all training X-rays are allowed to rotate up to 50 degrees, are feature-wise standardized using statistics from the entire set, can have their widths and heights shifted by up to 20\%, are allowed a shear intensity of up to .25 degrees in the counter-clockwise direction, as well as a zooming/de-zooming range of $\pm$10\%, a channel shifting range of .2, and can flip either horizontally or vertically. Missed pixels are filled through constant filling. This allows to also expand the number of training images. These allowed transformations for the training set are detailed in the Table \ref{tab:tabtrainingtransformations}.\\

\begin{figure}[h!]
     \centering
     \begin{subfigure}[t]{.23\textwidth}
         \centering
         \includegraphics[width = \textwidth]{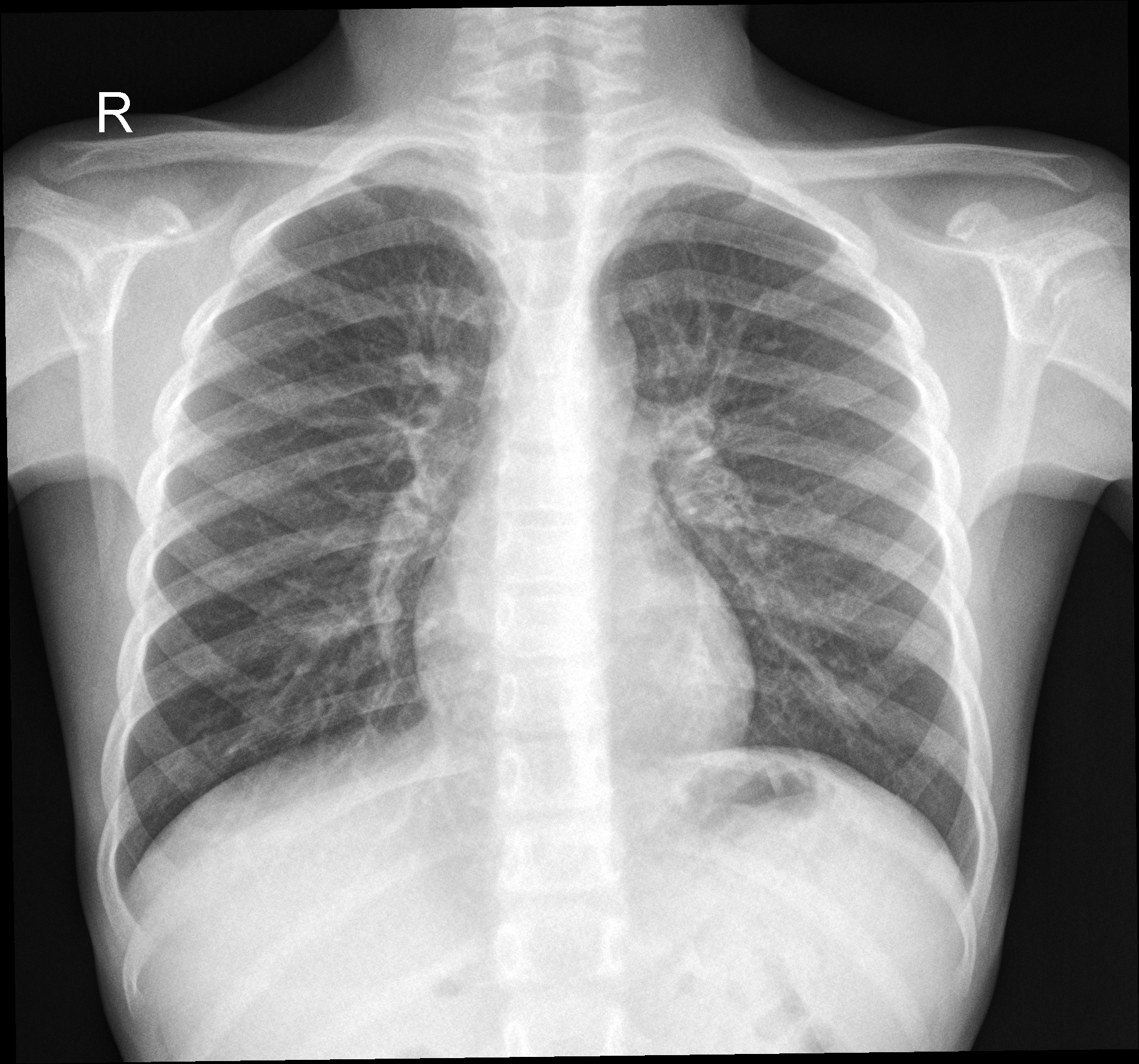}
         \caption{Lungs of a healthy patient.}
         \label{fig:ex_normal}
     \end{subfigure}
     \hfill
     \begin{subfigure}[t]{.23\textwidth}
         \centering
         \includegraphics[width = \textwidth]{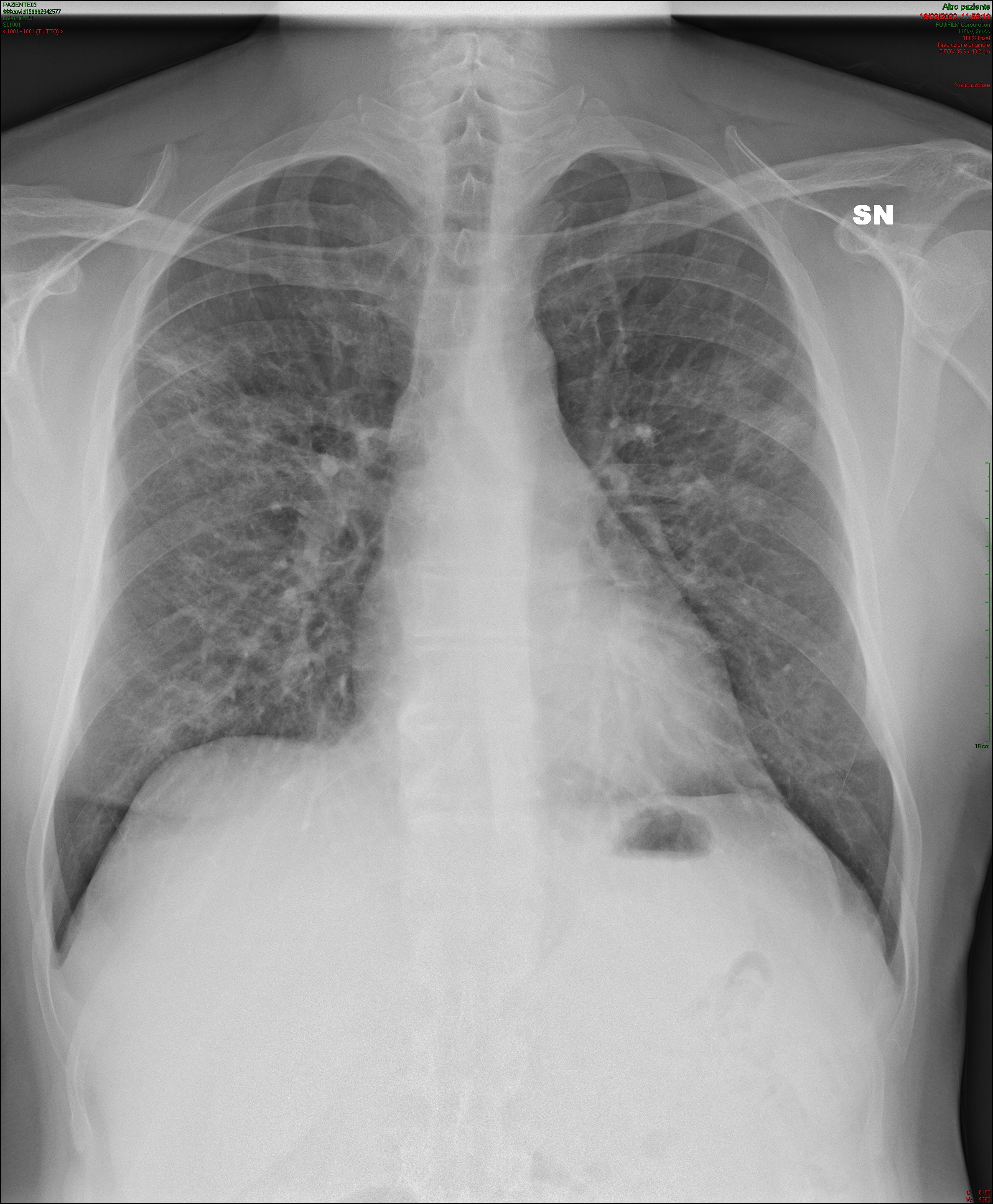}
         \caption{Lungs of a COVID-19 infected patient.}
         \label{fig:ex_covid19}
     \end{subfigure}
     \hfill
     \begin{subfigure}[t]{.23\textwidth}
         \centering
         \includegraphics[width = \textwidth]{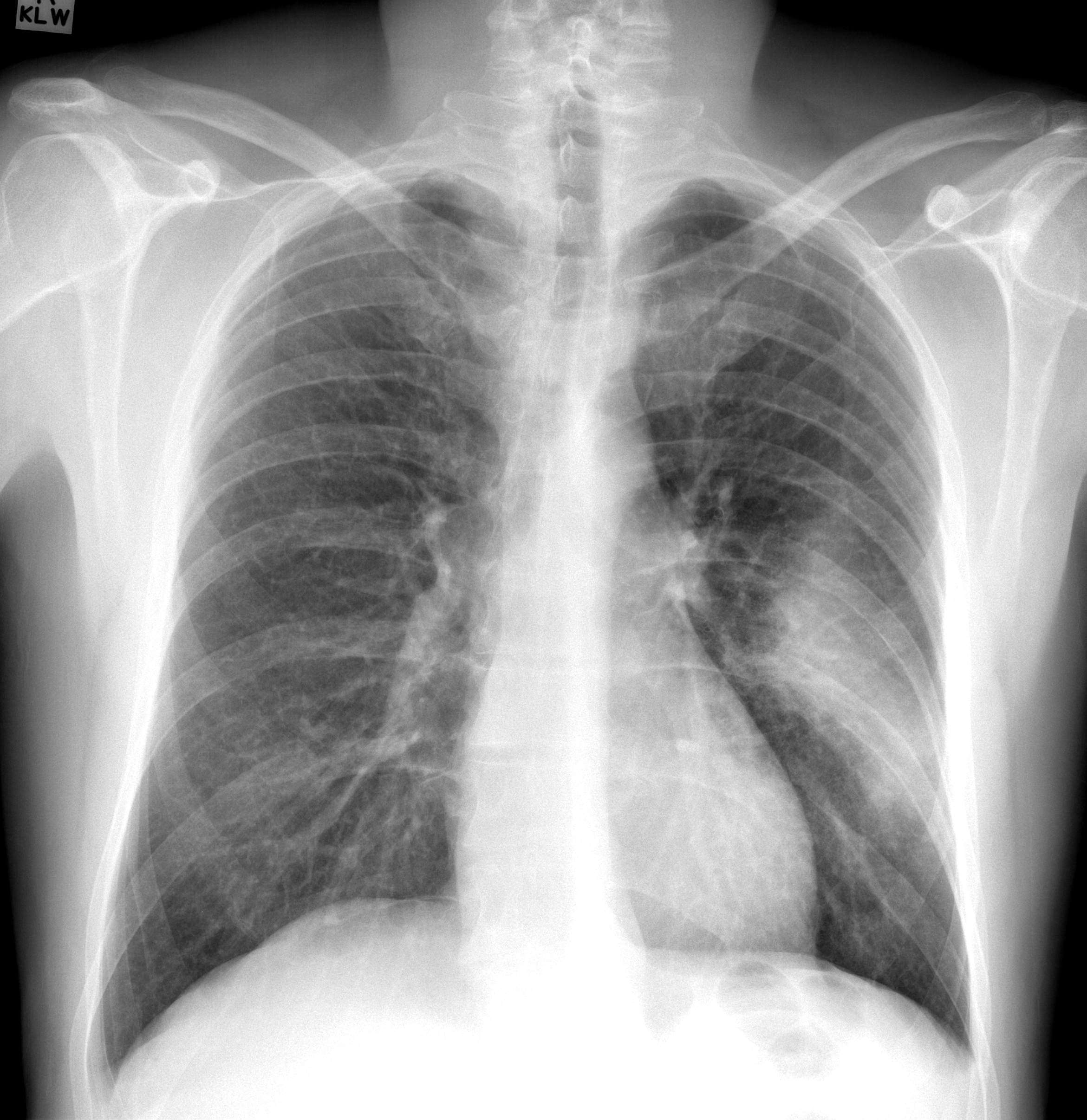}
         \caption{Lungs of a chlamydia infected patient.}
         \label{fig:ex_chlamydia}
     \end{subfigure}
     \hfill
     \begin{subfigure}[t]{.23\textwidth}
         \centering
         \includegraphics[width = \textwidth]{ex_ARDS.png}
         \caption{Lungs of a patient suffering ARDS.}
         \label{fig:ex_ARDS}
     \end{subfigure}
     ~
    \begin{subfigure}[t]{.98\textwidth}
        \centering
        \includegraphics[width = \linewidth]{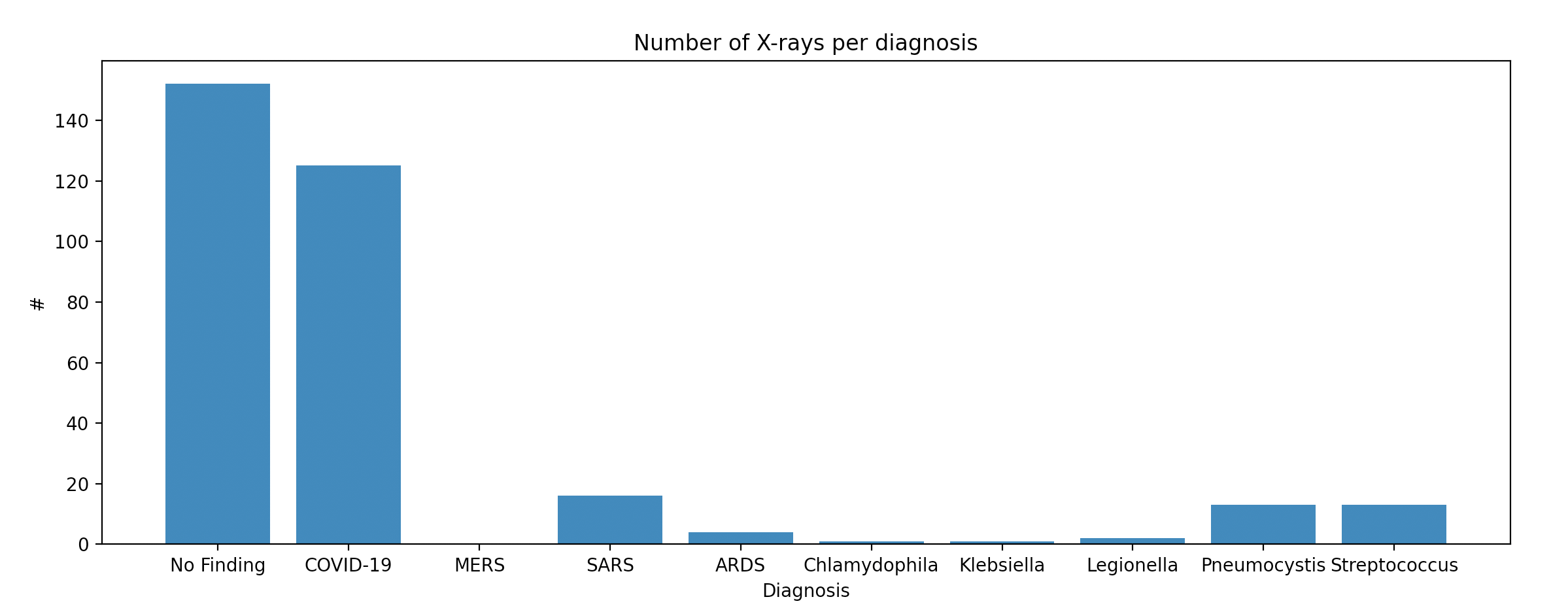}
        \caption{Final distribution of diagnoses by the healthcare professionals based on the PA chest X-rays in the combined data-sets.}
        \label{fig:data_category_dist}
    \end{subfigure}
        \caption{Examples of PA chest X-rays in the data-set in Figures \ref{fig:ex_normal} to \ref{fig:ex_ARDS}, and distribution of the diagnoses based on the PA chest X-rays in Figure \ref{fig:data_category_dist}.}
        \label{fig:examples}
\end{figure}

\begin{table}[h!]
    \centering
    \begin{tabular}{ll}
        \textbf{Feature} & \textbf{Comment} \\\hline
        \textbf{Resizing} & to $182\times182$ in RGB, resulting in $(182, 182, 3)$ arrays\\
        \textbf{Rotation} & up to $\pm50$ degrees\\
        \textbf{Standardization} & feature-wise through training set statistics\\
        \textbf{Width shift} & up to $\pm20\%$\\
        \textbf{Height shift} & up to $\pm20\%$\\
        \textbf{Shear intensity} & up to $.25$ degrees counter-clockwise\\
        \textbf{Zooming \& de-zooming} & up to $\pm10\%$\\
        \textbf{Channel shifting} & in the range of $20\%$\\
        \textbf{Horizontal \& vertical flip} & True
    \end{tabular}
    \caption{Possible transformations of the images in the training set before being introduced to the convolutional neural networks.}
    \label{tab:tabtrainingtransformations}
\end{table}

Figure \ref{fig:data_category_dist} shows the final distribution of illness diagnoses following the reading of the PA chest X-rays by the healthcare professionals.

\newpage
\section{VGG16}
The architecture of the first considered deep convolutional neural network follows the VGG16 architecture proposed by \citet{simonyan:2015}, and is detailed in Table \ref{tab:VGG16_architecture}. Once trained over 200 epochs on a data-set of over 14 million $224\times 224$ images belonging to 1,000 classes based on the Large Scale Visual Recognition Challenge 2014 (ILSVRC2014), this deep convolutional neural network achieved 92.7\% top-5 test accuracy. In this research, we kept the weights for pre-training following the ILSVRC2014. The optimizer is Adam \citep{kingma:2015} with categorical cross-entropy loss, learning rate 0.0001, and ran over 200 epochs.\\

\begin{table}[h!]
    \centering
    \begin{tabular}{llr}
        \textbf{Step type} & \textbf{Computations} & \textbf{Name}\\\hline\hline
        \multirow{2}{*}{First 2D-Convolutions} & 2D-Conv($3, 64$) & 2D-Conv\_111  \\
        & 2D-Conv($3, 64$) & 2D-Conv\_112 \\\hline
        First 2D-Maximum pooling & MaxPool($2, 2$) & Pool\_11\\\hline
        \multirow{2}{*}{Second 2D-Convolutions} & 2D-Conv($3, 128$) & 2D-Conv\_121 \\
        & 2D-Conv($3, 128$) & 2D-Conv\_122 \\\hline
        Second 2D-Maximum pooling & MaxPool($2, 2$) & Pool\_12 \\\hline
        \multirow{3}{*}{Third 2D-Convolutions} & 2D-Conv($3, 256$) & 2D-Conv\_131 \\
        & 2D-Conv($3, 256$) & 2D-Conv\_132 \\
        & 2D-Conv($3, 256$) & 2D-Conv\_133\\ \hline
        Third 2D-Maximum pooling & MaxPool($2, 2$) & Pool\_13\\\hline
        \multirow{3}{*}{Fourth 2D-Convolutions} & 2D-Conv($3, 512$) & 2D-Conv\_141 \\
        & 2D-Conv($3, 512$) & 2D-Conv\_142 \\
        & 2D-Conv($3, 512$) & 2D-Conv\_143 \\ \hline
        Fourth 2D-Maximum pooling & MaxPool($2, 2$) & Pool\_14\\\hline
        \multirow{3}{*}{Fifth 2D-Convolutions} & 2D-Conv($3, 512$) & 2D-Conv\_151 \\
        & 2D-Conv($3, 512$) & 2D-Conv\_152 \\
        & 2D-Conv($3, 512$) & 2D-Conv\_153 \\ \hline
        Fifth 2D-Maximum pooling & MaxPool($2, 2$) & Pool\_15\\\hline
        \multirow{7}{*}{Flat portion} & Flatten & Flat\_11 \\ 
        & Layer($4096$) & Layer\_11\\
        & Layer($4096$) & Layer\_12\\
        & Layer($1000$) & Layer\_13 \\
        & Batch Normalization & Norm\_11\\
        & Layer($256$) & Layer\_14\\
        & DropOut($p$) & Drop\_11 \\\hline
        Output& Output(softmax) & Out\_1\\\hline
    \end{tabular}
    \caption{VGG16 configuration \citep{simonyan:2015} with an added flat portion that has some drop-out rate $p$. The activation function is Rectified Linear Unit (ReLU) throughout. The total number of parameters for this architecture is 17,994,563 for $182\times182$ images.}
    \label{tab:VGG16_architecture}
\end{table}

We perform the Stratified 5-Fold cross-validation of the VGG16 with 30\% final drop-out, and the results are as shown in Table \ref{tab:VGG16_results}.\\

\begin{table}[h!]
    \centering
    \begin{tabular}{lccc}
        \multicolumn{4}{c}{\textsc{Internal / Training Set}}\\\hline\hline
        \textbf{Measure} & \textbf{LHS 95\% CI} & \textbf{Value} & \textbf{RHS 95\% CI} \\\hline
        \textbf{Loss} & -0.135 & 0.105 & 0.339\\
        \textbf{Accuracy} & 0.905 & 0.939 & 0.973\\
        \textbf{Flat AUC} & 0.971 & 0.974 & 0.976\\
        \textbf{COVID-19 Recall} & 0.858 & 0.877 & 0.897\\
        \textbf{No Finding Recall} & 0.960 & 0.968 & 0.976\\
        \textbf{Other Pneumonia Recall} & 0.500 & 0.534 & 0.568\\\hline
        \multicolumn{4}{c}{\textsc{External / Testing Set}}\\\hline\hline
        \textbf{Measure} & \textbf{LHS 95\% CI} & \textbf{Value} & \textbf{RHS 95\% CI} \\\hline
        \textbf{Loss} & -0.324 & 0.337 & 0.998\\
        \textbf{Accuracy} & 0.706 & 0.841 & 0.976\\
        \textbf{Flat AUC} & 0.970 & 0.974 & 0.977\\
        \textbf{COVID-19 Recall} & 0.858 & 0.877 & 0.897\\
        \textbf{No Finding Recall} & 0.960 & 0.968 & 0.976\\
        \textbf{Other Pneumonia Recall} & 0.500 & 0.534 & 0.568\\
    \end{tabular}
    \caption{Cross-validated performance measures (rounded to the thousandth) of the VGG16 neural network outlined in Table \ref{tab:VGG16_architecture} through Stratified 5-Fold and 200 epochs, learning rate of 0.0001. The 95\% confidence interval were computed through the $t$-distribution with 4 degrees of freedom.}
    \label{tab:VGG16_results}
\end{table}

In this case, we find satisfactory accuracies for the training and testing set, of respectively $94$\% and $84$\%, although the 95\% confidence is quite large for the latter. Interestingly, the COVID-19 sensitivity remains high with short-range confidence intervals. A $88$\% COVID-19 sensitivity is similar to \citet{wang:2020} for the training set, but significantly higher for the testing set. In clinical terms, around 88\% of those diagnosed or tested COVID-19 positive are identified as such by the neural network outlined in Table \ref{tab:VGG16_architecture}. Similarly, we found that the No Finding sensitivity, i.e., the percentage of lung-healthy patients identified as such, is of 97\% for both the training and testing set.\\

On the other hand, the sensitivity for pneumonia diseases other than COVID-19 is poor (at $53$\% for both the training and testing sets), meaning the neural network has difficulties differentiating COVID-19 and other pneumoniae. For such differentiation, \citet{wang:2020} found a sensitivity of $67$\%, which is better, although they only used two classes, namely COVID-19 and Other Pneumonia. It is possible that this improved sensitivity is due to the fact that \citet{wang:2020} trained their neural networks on CT images, and not on PA X-rays, which tend to better differentiate pathologies. Furthermore, there is an under-representation of such label in our data-set, although we consider that it could be representative of data-sets found in hospital during a COVID-19 epidemic.\\

As we are dealing with a multi-class problem, the Flat AUC (Area Under the ROC Curve) is not equivalent to the standard binary AUC. We here flattened the data into a single label before AUC computation. In which case, each label-prediction pair is treated as an individual data point. Therefore, this measure should not be very reliable, although does procure some information on the "overall" ROC of the neural network. It is found to be of $97.4$\% for both the training and testing data-sets. Due to the construction of this performance measure, it is better to compare it to other networks and may not carry significance of the network alone -- which is what we will do in the sections below.

\section{InceptionResNetV2 \& InceptionV3}

\begin{figure}[h!]
    \centering
    \includegraphics[width = 1\linewidth]{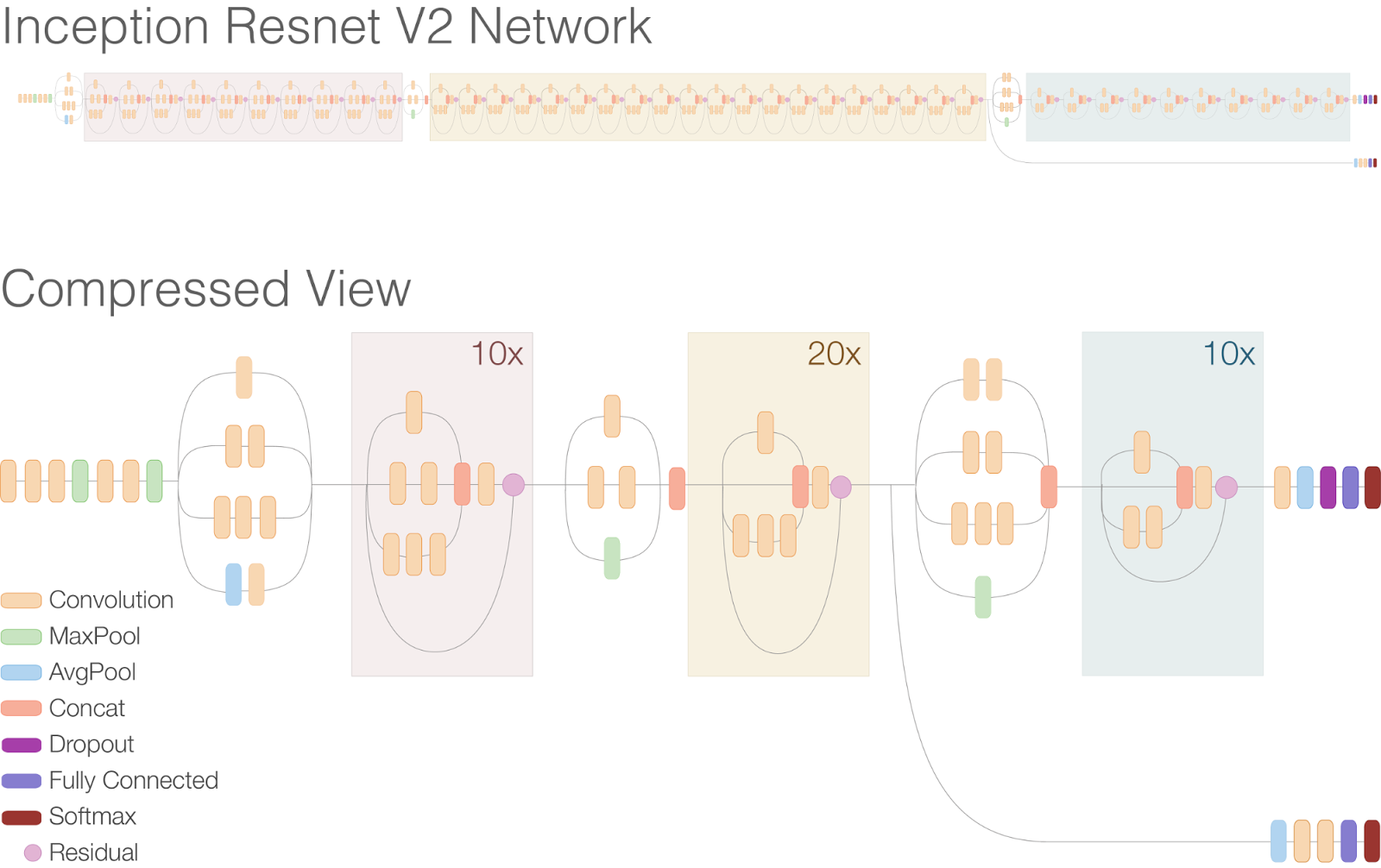}
    \caption{Description diagram of the architecture of InceptionResNetV2, as detailed in \citet{szegedy:2016}. On the $182\times182$ data-set, this neural network has 54,737,123 parameters.}
    \label{fig:InceptionResNetV2_architecture}
\end{figure}

The second architecture considered is a very deep convolutional neural network named InceptionResNetV2 proposed by \citet{szegedy:2016}, and it is described in the diagram shown in Figure \ref{fig:InceptionResNetV2_architecture}. Once trained over 200 epochs on the ImageNet data-set, this very deep convolutional neural network achieved 95.3\% top-5 test accuracy. In this research, we kept the weights for pre-training after the above training, we use an Adam optimizer \citep{kingma:2015} with learning rate 0.0001, categorical cross-entropy loss, and 200 epochs. In the architecture, max-pooling was performed for feature extraction.\\

The cross-validated performance measures, obtained through Stratified 5-Fold, for the InceptionResNetV2 with final 30\% drop-out are shown in Table \ref{tab:IRNV2_results}.\\

\begin{table}[h!]
    \centering
    \begin{tabular}{lccc}
        \multicolumn{4}{c}{\textsc{Internal / Training Set}}\\\hline\hline
        \textbf{Measure} & \textbf{LHS 95\% CI} & \textbf{Value} & \textbf{RHS 95\% CI} \\\hline
        \textbf{Loss} & 0.073 & 0.977 & 1.881\\
        \textbf{Accuracy} & 0.520 & 0.616 & 0.712\\
        \textbf{Flat AUC} & 0.906 & 0.918 & 0.931\\
        \textbf{COVID-19 Recall} & 0.673 & 0.708  & 0.744\\
        \textbf{No Finding Recall} & 0.911 & 0.918 & 0.925\\
        \textbf{Other Pneumonia Recall} & 0.351 & 0.433 & 0.514\\\hline
        \multicolumn{4}{c}{\textsc{External / Testing Set}}\\\hline\hline
        \textbf{Measure} & \textbf{LHS 95\% CI} & \textbf{Value} & \textbf{RHS 95\% CI} \\\hline
        \textbf{Loss} & 0.488 & 1.411 & 2.335\\
        \textbf{Accuracy} & 0.457 & 0.612 & 0.765\\
        \textbf{Flat AUC} & 0.907 & 0.919 & 0.932\\
        \textbf{COVID-19 Recall} & 0.675 & 0.710 & 0.745\\
        \textbf{No Finding Recall} & 0.911 & 0.918 & 0.925\\
        \textbf{Other Pneumonia Recall} & 0.353 & 0.433 & 0.513\\
    \end{tabular}
    \caption{Cross-validated performance measures (rounded to the thousandth) of the InceptionResNetV2-like neural network outlined in Figure \ref{fig:InceptionResNetV2_architecture} through Stratified 5-Fold and 200 epochs, learning rate of 0.0001. The 95\% confidence interval were computed through the $t$-distribution with 4 degrees of freedom.}
    \label{tab:IRNV2_results}
\end{table}

The performance results for InceptionResNetV2 tend here to be quite disappointing, mainly because this architecture includes many batch normalizations \citep{ioffe:2015}. During model training, the batches are normalized by their mean and variance, but in the testing phase, the batches are normalized with respect to the changing average of observed mean and variance, leading to lower accuracies for the type of data-set we are considering.\\

In particular, we found COVID-19 sensitivities for the InceptionResNetV2 network of $70.8(-3.5,+3.6)$\% and $71.0(\pm3.5)$\% for the internal and external validations respectively. In medical terms, this is considered relatively poor, and seems indeed low even if we take into account errors from RT-PCR testing or seroprevalence studies as showcased in \citet{xu:2020}. Furthermore, for the accuracies and COVID-19 sensitivities, the 95\% confidence interval implied by the Stratified 5-Fold cross-validation is wide, implying greater variance. Hence, it is likely that training and testing on different but similar data-sets could lead to very different performance results. \\

Similarly, we can compare the Flat AUC value with the VGG16-like network trained as in the previous section. According to the measures gathered, the VGG16-like neural network improved the internal and external Flat AUCs by approximately 6\% compared to the InceptionResNetV2-like neural network. Hence, we believe that the VGG16-like neural network is more appropriate for the data-set we are studying compared to the InceptionResNetV2-like neural network.\\

In the case of the InceptionV3 convolutional neural network \citep{szegedy:2015}, with the same optimizer, we find similar albeit slightly enhanced performance results compared to InceptionResNetV2, as seen in Table \ref{tab:InceptionV3_results}.

\begin{table}[h!]
    \centering
    \begin{tabular}{lccc}
        \multicolumn{4}{c}{\textsc{Internal / Training Set}}\\\hline\hline
        \textbf{Measure} & \textbf{LHS 95\% CI} & \textbf{Value} & \textbf{RHS 95\% CI} \\\hline
        \textbf{Loss} & -2.913 & 2.676 & 8.266\\
        \textbf{Accuracy} & 0.529 & 0.705 & 0.880\\
        \textbf{Flat AUC} & 0.908 & 0.931 & 0.954\\
        \textbf{COVID-19 Recall} & 0.689 & 0.779 & 0.870\\
        \textbf{No Finding Recall} & 0.928 & 0.941 & 0.953\\
        \textbf{Other Pneumonia Recall} & 0.421 & 0.491 & 0.561\\\hline
        \multicolumn{4}{c}{\textsc{External / Testing Set}}\\\hline\hline
        \textbf{Measure} & \textbf{LHS 95\% CI} & \textbf{Value} & \textbf{RHS 95\% CI} \\\hline
        \textbf{Loss} & -0.757 & 3.477 & 7.711\\
        \textbf{Accuracy} & 0.477 & 0.691 & 0.905\\
        \textbf{Flat AUC} & 0.909 & 0.931 & 0.954\\
        \textbf{COVID-19 Recall} & 0.691 & 0.780 & 0.869\\
        \textbf{No Finding Recall} & 0.928 & 0.941 & 0.953\\
        \textbf{Other Pneumonia Recall} & 0.422 & 0.491 & 0.560\\
    \end{tabular}
    \caption{Cross-validated performance measures (rounded to the thousandth) of the InceptionV3-like neural network through Stratified 5-Fold and 200 epochs, learning rate of 0.0001. The 95\% confidence interval were computed through the $t$-distribution with 4 degrees of freedom.}
    \label{tab:InceptionV3_results}
\end{table}

\section{VGG19}

Due to the earlier success of VGG16, we consider the VGG19 deep convolutional neural network architecture as showcased in \citet{simonyan:2015}, and again with 200 epochs, the Adam optimizer \citep{kingma:2015}, a learning rate of 0.0001, and categorical cross-entropy loss. VGG19 is similar to VGG16, instead we add the following convolutions \citep{simonyan:2015}: 
\begin{itemize}
    \item 2D-Conv(3, 256) after 2D\_Conv\_133 in Table \ref{tab:VGG16_architecture},
    \item 2D-Conv(3, 512) after 2D\_Conv\_143 in Table \ref{tab:VGG16_architecture},
    \item and 2D-Conv(3, 512) after 2D\_Conv\_153 in Table \ref{tab:VGG16_architecture}.
\end{itemize}

\begin{table}[h!]
    \centering
    \begin{tabular}{lccc}
        \multicolumn{4}{c}{\textsc{Internal / Training Set}}\\\hline\hline
        \textbf{Measure} & \textbf{LHS 95\% CI} & \textbf{Value} & \textbf{RHS 95\% CI} \\\hline
        \textbf{Loss} & -0.056 & 0.082 & 0.220\\
        \textbf{Accuracy} & 0.902 & 0.927 & 0.953\\
        \textbf{Flat AUC} & 0.967 & 0.969 & 0.971\\
        \textbf{COVID-19 Recall} & 0.843 & 0.861 & 0.879\\
        \textbf{No Finding Recall} & 0.960 & 0.964 & 0.969\\
        \textbf{Other Pneumonia Recall} & 0.428 & 0.480 &  0.531\\\hline
        \multicolumn{4}{c}{\textsc{External / Testing Set}}\\\hline\hline
        \textbf{Measure} & \textbf{LHS 95\% CI} & \textbf{Value} & \textbf{RHS 95\% CI} \\\hline
        \textbf{Loss} & -0.527 & 0.485 & 1.498\\
        \textbf{Accuracy} & 0.702 & 0.820 & 0.937\\
        \textbf{Flat AUC} & 0.967 & 0.969 & 0.970\\
        \textbf{COVID-19 Recall} & 0.843 & 0.861 & 0.879\\
        \textbf{No Finding Recall} & 0.960 & 0.964 & 0.968\\
        \textbf{Other Pneumonia Recall} & 0.427 & 0.479 & 0.531\\
    \end{tabular}
    \caption{Cross-validated performance measures (rounded to the thousandth) of the VGG19-like neural network through Stratified 5-Fold and 200 epochs, learning rate of 0.0001. The 95\% confidence interval were computed through the $t$-distribution with 4 degrees of freedom.}
    \label{tab:VGG19_results}
\end{table}

The performance measures seen in Table \ref{tab:VGG19_results} for the VGG19-like network are very similar than those gathered for the VGG16-like network (Table \ref{tab:VGG16_results}) due to the very similar architecture. In particular, we find that the added convolutions do not increase significantly the performance measures, although we see that the 95\% confidence intervals are shrunk. Furthermore, due to these added convolutions, VGG19 is slightly more costly than VGG16 despite no big improvement in the measures.

\newpage
\section{Conclusion}

\begin{figure}[h!]
     \centering
     \begin{subfigure}[t]{.75\textwidth}
         \centering
         \includegraphics[width = \textwidth]{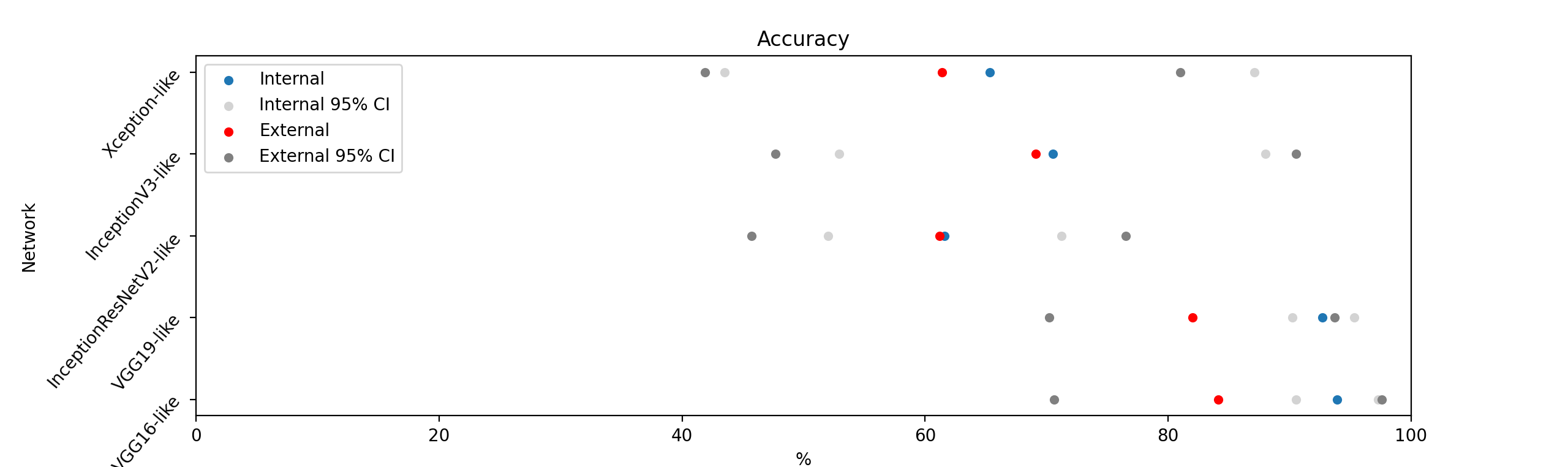}
         \label{fig:acc_con}
     \end{subfigure}
     ~
     \begin{subfigure}[t]{.75\textwidth}
         \centering
         \includegraphics[width = \textwidth]{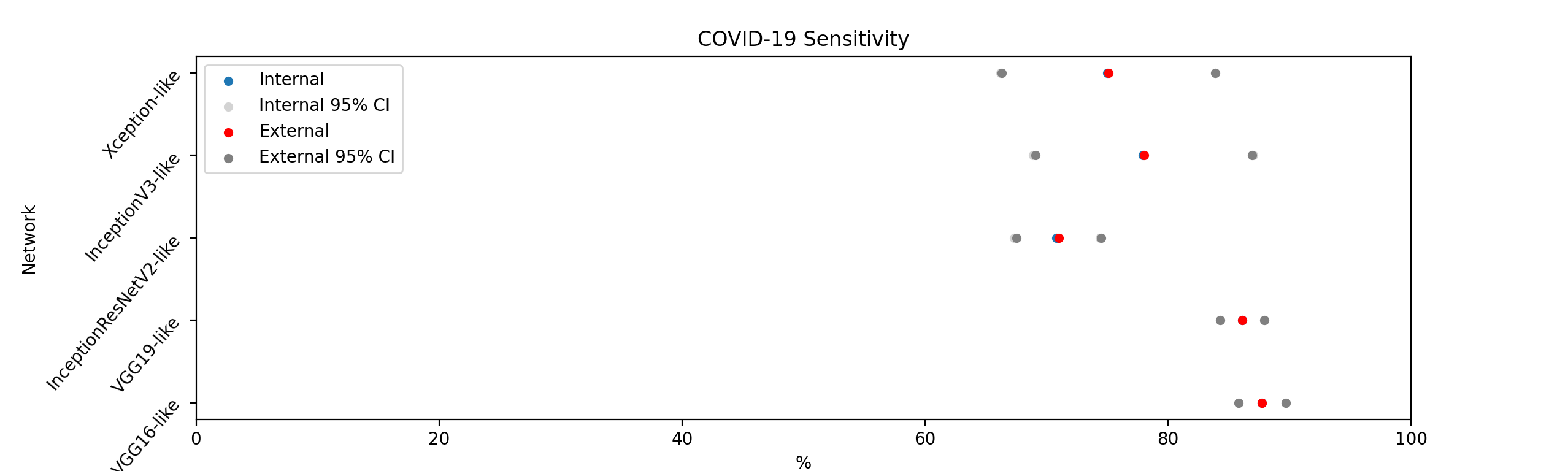}
         \label{fig:covid_con}
     \end{subfigure}
     ~
     \begin{subfigure}[t]{.75\textwidth}
         \centering
         \includegraphics[width = \textwidth]{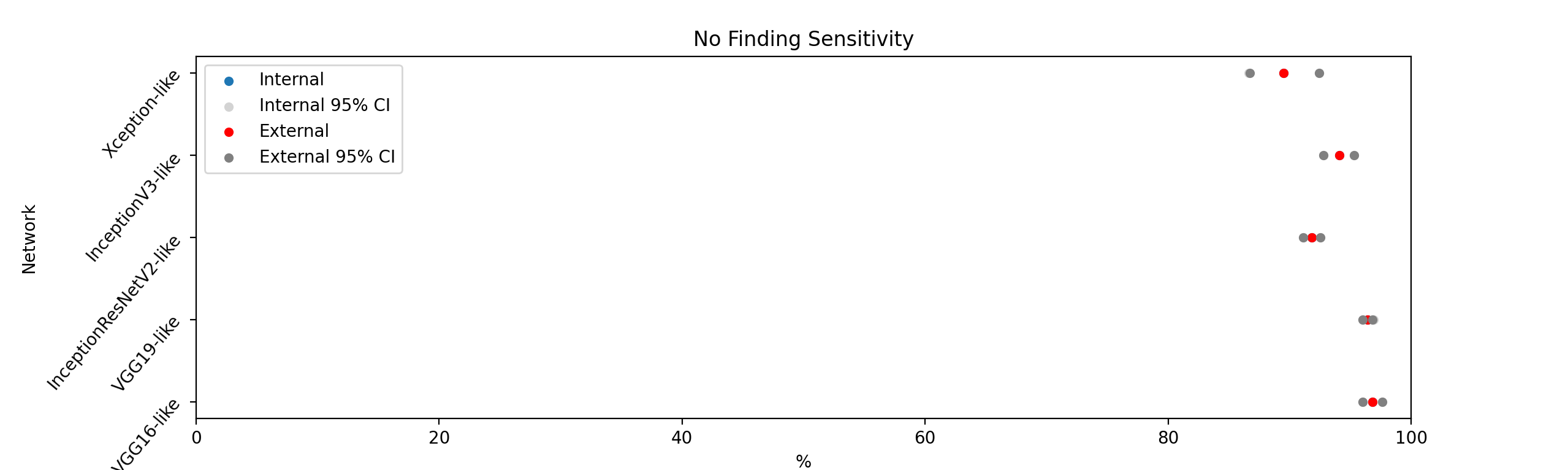}
         \label{fig:nof_con}
     \end{subfigure}
     ~
     \begin{subfigure}[t]{.75\textwidth}
         \centering
         \includegraphics[width = \textwidth]{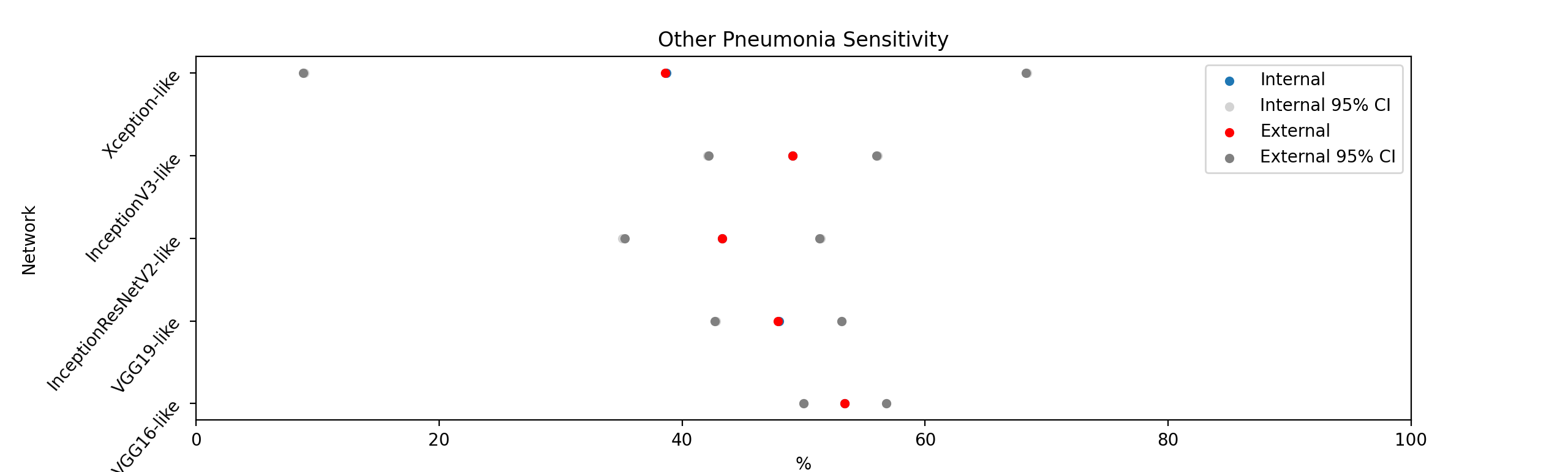}
         \label{fig:otherpneu_con}
     \end{subfigure}
     ~
    \begin{subfigure}[t]{.75\textwidth}
        \centering
        \includegraphics[width = \linewidth]{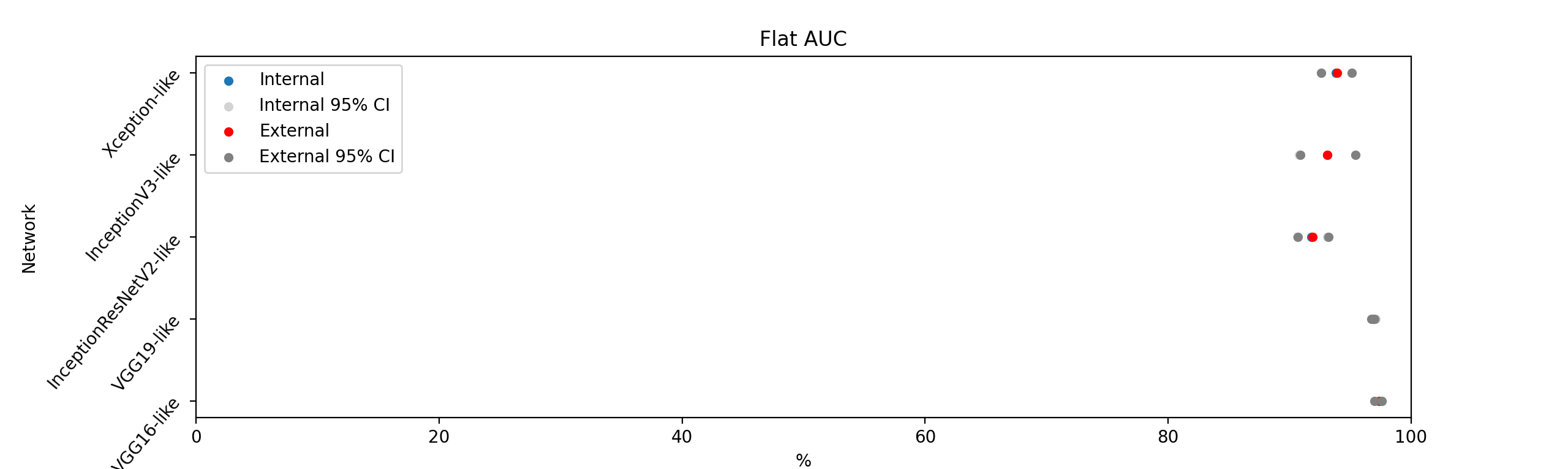}
        \label{fig:flatauc_con}
    \end{subfigure}
        \caption{Performance measures of various neural networks on the data-set of PA chest X-rays.}
        \label{fig:con}
\end{figure}

A summary table of some of the performance measures for the neural networks all trained over 200 epochs are shown in Table \ref{tab:conclusion_mf}. Some of these measures are further shown in Figure \ref{fig:con}. \\

As we can see from Table \ref{tab:conclusion_mf} and Figure \ref{fig:con}, VGG16 performed best on all measures, although VGG19 had a similar performance with lower variability. In particular, VGG16 had Stratified 5-Fold cross-validated accuracies of $93.9(\pm3.4)$\% for the internal data-set and of $84.1(\pm13.5)$\% for the external data-set. Note that we have run all the training sessions with 200 epochs for comparability between the neural networks, and therefore this difference between the internal and external accuracies might be explained by over-fitting. Furthermore, the COVID-19 sensitivity of the VGG16 neural network is of $87.7(-1.9,2)$\% for both the internal and external data-sets, suggesting efficiency of that network in correctly identifying true COVID-19 positives. Idem for the No Finding sensitivity which is at $96.8(\pm0.8)$\%. \\

{ 
\renewcommand{\arraystretch}{1.82}
\begin{table}[h!]
    \centering
    \begin{tabular}{l|ccc|ccc}
        - & \multicolumn{3}{c|}{\textsc{Internal / Training Set}} & \multicolumn{3}{c}{\textsc{External / Testing Set}}\\\hline\hline
        \textsc{Net} & \textbf{Accuracy} & \textbf{\#1 Recall} & \textbf{\#2 Recall} & \textbf{Accuracy} & \textbf{\#1 Recall} & \textbf{\#2 Recall} \\\hline
        (1) & $93.9_{-3.4}^{+3.4}$\% & $87.7_{-1.9}^{+2.0}$\% & $96.8_{-0.8}^{+0.8}$\% & $84.1_{-13.5}^{+13.5}$\% & $87.7_{-1.9}^{+2.0}$\% & $96.8_{-0.8}^{+0.8}$\%\\
        (2) & $92.7_{-2.5}^{+2.6}$\% & $86.1_{-1.8}^{+1.8}$\% & $96.4_{-0.4}^{+0.5}$\% & $82.0_{-11.8}^{+11.7}$\% & $86.1_{-1.8}^{+1.8}$\% & $96.4_{-0.4}^{+0.4}$\%\\
        (3) & $61.6_{-9.6}^{+9.6}$\% & $70.8_{-3.5}^{+3.6}$\% & $91.8_{-0.7}^{+0.7}$\% & $61.2_{-15.5}^{+15.3}$\% & $71.0_{-3.5}^{+3.5}$\% & $91.8_{-0.7}^{+0.7}$\% \\
        (4) & $70.5_{-17.6}^{+17.5}$\% & $77.9_{-9.0}^{+9.1}$\% & $94.1_{-1.3}^{+1.2}$\% & $69.1_{-21.4}^{+21.4}$\% & $78.0_{-8.9}^{+8.9}$\% & $94.1_{-1.3}^{+1.2}\%$\\
        (5) & $65.3_{-21.8}^{+21.8}$\% & $75.0_{-8.8}^{+8.9}$\% & $89.5_{-2.9}^{+2.9}$\% & $61.4_{-19.5}^{+19.6}$\% & $75.1_{-8.8}^{+8.8}$\% & $89.5_{-2.8}^{+2.9}$\% \\
    \end{tabular}
    \caption{Cross-validated (Stratified 5-Fold) internal and external measures for each of the models considered, rounded to the closest thousandth. Here, \#1 Recall refers to the sensitivity of the neural network on the COVID-19 categorical variable, and \#2 Recall the sensitivity on the No Finding categorical variable. Networks: (1) VGG16 \citep{simonyan:2015}, (2) VGG19 \citep{simonyan:2015}, (3) InceptionResNetV2 \citep{szegedy:2016}, (4) InceptionV3 \citep{szegedy:2016}, and (5) Xception \citep{chollet:2017}, all with a 30\% final drop-out.}
    \label{tab:conclusion_mf}
\end{table}
}

One topic of research that could be pursued following this paper is the creation of a specific deep neural network tailored for the data-set made of PA chest X-rays. This research would be dependent on the easy access of a greater database of PA chest X-rays COVID-19 positive patients (through CT diagnosis, RT-PCR or seroprevalence testing), as well as the computational capabilities of large-scale training of very deep convolutional networks. Furthermore, such tailored network could make it harder to share it to hospitals, and as a consequence making it less prone to front-line applications.\\

\newpage

\section{Limits of this paper}

The biggest limit to this research is by far the data-set. The data-set collected was small after balancing for the number of COVID-19 positive cases, and still was quite unbalanced as the class "Other Pneumonia" was under-represented. Furthermore, the data includes PA chest X-rays from a multitude of hospitals, which could have lowered the accuracies and sensitivities. \\

It is also clear that the predictions on a PA chest X-ray given by the trained deep neural networks cannot be relied upon alone. Its best use is in conjunction of clinical tests (such as pathogen-RT-PCR or seroprevalence testing), or professional diagnosis based on other clearer medical imaging methods such as CT scans.

\section{For hospitals}

Certified hospitals can request the code by contacting the author and / or Imperial College London at \href{mailto:pierre.moutounet-cartan17@imperial.ac.uk}{pierre.moutounet-cartan17@imperial.ac.uk}, as well as for explanation of how to use it with their own data-set (or a collective data-set between "sister" hospitals).\\

There are two main goals to this research to help hospitals:
\begin{itemize}
    \item in the medium to long-term, being able to diagnose through machine learning tools possible pneumonia, and if detected, whether it is linked to a coronavirus infection or other, allowing the detection of new possible coronavirus foyers after the end of possible "stop-and-go" lockdowns as expected by \citet{ferguson:2020} until a vaccine is found and widespread;
    \item in the short-term, it can allow practitioners to compare the diagnosis from the deep convolutional neural networks in this paper with possible RT-PCR testing results -- if clashing, a chest CT could be done as they are more accurate in showing COVID-19 pneumonia as shown by \citet{ai:2020, wang:2020, xu:2020}.
\end{itemize}

This research focuses on PA chest X-rays because, although they can be less accurate to diagnose COVID-19 than CT imaging, they are more common in hospitals and relatively cheaper to perform. Hence, training neural networks on PA chest X-rays could lead to greater generalization and application in hospitals around the world, and CT imaging would only be used for cases where patients are pathogen-confirmed COVID-19 through RT-PCR testing but the AI diagnoses No Finding, or when the AI diagnoses the patients as highly-likely COVID-19 but the RT-PCR tests are negative.

\section{Conflicts of interest}

The author(s) note no known conflict of interest by undertaking this research.

\newpage

\vskip .2in
\bibliography{sample}

\end{document}